\documentclass[aps,twocolumn,amsmath,amssymb,preprintnumbers,superscriptaddress,prl,floatfix]{revtex4-2}

\usepackage{comment}
\usepackage[version=4]{mhchem}
\usepackage[utf8]{inputenc}
\usepackage{newtxtext}
\usepackage[upint]{newtxmath}
\usepackage{microtype}
\usepackage{textcomp}
\usepackage{dsfont}
\usepackage{eucal}
\usepackage{siunitx}
\usepackage{soul}

\usepackage{enumerate}
\usepackage{amsfonts}
\usepackage{color}
\usepackage{soul}

\usepackage{todonotes}
\presetkeys%
    {todonotes}%
    {inline}{}

\usepackage{graphicx}

\usepackage[colorlinks,allcolors=blue]{hyperref}
\usepackage[capitalize]{cleveref} 
\usepackage{cleveref}

\newcommand{\e}[1]{\mathrm{e}^{#1}}

\newcommand{\vecQ}{\boldsymbol{Q}}
\newcommand{\vecR}{\boldsymbol{R}}
\newcommand{\vk}{\boldsymbol{k}}

\newcommand{\vecx}{\boldsymbol{x}}
\newcommand{\vecz}{\boldsymbol{z}}

\newcommand{\ii}{\text{i}}

\newcommand{\vecr}{\boldsymbol{r}}      

\definecolor{DarkBlue}{rgb}{0,0,0.80}
\definecolor{DarkRed}{rgb}{0.80,0,0}
\definecolor{Purple}{rgb}{0.55,0,0.55}
\definecolor{Purple}{rgb}{0,0,0.8}

\newcommand{\Chi}[1]{\textcolor{DarkBlue}{#1}}

\let\epsilon\varepsilon

\begin{document}

\title{Supercurrent-induced spin switching via indirect exchange interaction}
\author{Chi Sun}
\affiliation{Center for Quantum Spintronics, Department of Physics, Norwegian \\ University of Science and Technology, NO-7491 Trondheim, Norway}
\author{Jacob Linder}
\affiliation{Center for Quantum Spintronics, Department of Physics, Norwegian \\ University of Science and Technology, NO-7491 Trondheim, Norway}\date{\today}
\begin{abstract}
Localized spins of single atoms adsorbed on surfaces have been proposed as building blocks for spintronics and quantum computation devices. However, identifying a way to achieve current-induced switching of spins with very low dissipation is an outstanding challenge with regard to practical applications. Here, we show that the indirect exchange interaction between spin impurities can be controlled by a dissipationless supercurrent.
All that is required is a conventional superconductor and two spin impurities placed on its surface. No triplet Cooper pairs or exotic material choices are needed. This finding provides a new and accessible way to achieve the long-standing goal of supercurrent-induced spin switching.
\end{abstract}

\maketitle
\textit{Introduction --} Electrical manipulation of spin or magnetization is crucial in the development of spintronics devices and technologies for data storage and computation \cite{brataas_natmat_12,Shi2019Oct,Yamada2007Apr}. Current-induced magnetization switching is presently used in magnetic random access memory through spin-transfer torque \cite{chun_jssc_13, tudu_vacuum_17}. However, the inclusion of electric current inevitably involves Joule heating and therefore high energy dissipation. An important objective is therefore to identify a way to electrically switch the directions of spins with very low dissipation. In spintronics, major efforts have been devoted to optimizing the choice of materials and hybrid structures \cite{Li2019Aug,Fukami2016May,Safeer2016Feb,Fan2014Jul,Sun2019Sep} in order to reduce the power consumption for switching, making it comparable to that in present semiconductor field-effect transistors \cite{e_field_compare}.

At low temperatures, an obvious candidate for achieving low-dissipation electric control over magnetism is superconducting materials due to their ability to host dissipationless supercurrents. Combining superconductivity and spintronics \cite{linder_nphys_15} offers possibilities to achieve supercurrent-induced magnetization dynamics, by which Joule heating and dissipation can be minimized. To achieve this, several theory papers have proposed to utilize spin-polarized triplet supercurrents \cite{eschrig_rpp_15}, which have been experimentenlly verified in superconductor (SC)/ferromagnet (FM) Josephson junctions \cite{keizer_nature_06, robinson_science_10, khaire_prl_10, super_CrO2}. It has also been theoretically shown that triplet supercurrents can
induce spin-transfer torque switching \cite{super_stt1,super_stt2, Linder2011Jan} and magnetization
dynamics \cite{buzdin_prl_08, teber_prb_10, teber_prb_11,  kulagina_prb_14, hals_prb_16, rabinovich_prl_19, bobkova_prb_20}. However, there exists no experimental observation of supercurrent-induced torque or magnetization dynamics. Part of the challenge lies within the complexity of the appropriate fabrication of the SC/FM multilayered structures, in which the SC/FM interface plays an essential role to create the triplet Cooper pairs for spin-polarized supercurrent.

\begin{figure}[t!]
    \centering
    \includegraphics[width = 0.48\textwidth]{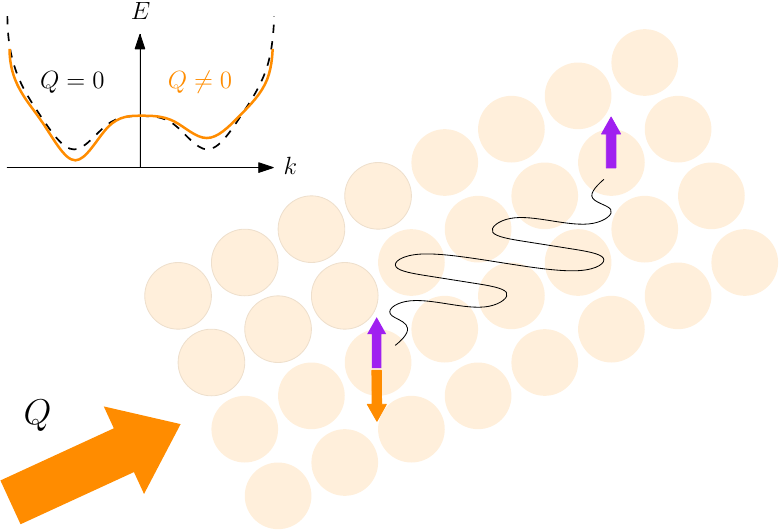}
    \caption{(Color online) Two impurity spins (purple small arrows) are coupled via the RKKY interaction (wavy black line) mediated by conduction band quasiparticles in the superconducting state. The picture shows a scenario where a parallell spin orientation is energetically preferred. When a supercurrent (large orange arrow) is applied, giving the Cooper pairs a finite momentum $Q$,  the quasiparticle bands become asymmetric in momentum $k$ (upper part of the plot), due to the broken parity symmetry. This causes a change in the RKKY interaction which can now favor the opposite spin orientation, in this case antiparallell (small orange arrow). In this way, the supercurrent induces spin switching.} 
    \label{fig:model}
\end{figure}

In this work, we theoretically demonstrate a new and conceptually simple way in which the goal of spin switching via singlet supercurrents can be achieved. We consider a conventional SC and two spin impurities placed on its surface (see Fig. \ref{fig:model}). Without the requirement of triplet Cooper pair or exotic material choices, this is a drastically simpler setup than previous studies, theoretical and experimental, that have considered magnetization dynamics in the superconducting state. By investigating the indirect exchange interaction between the two spin impurities, also known as the Ruderman-Kittel-Kasuya-Yosida (RKKY) interaction \cite{ruderman_pr_54, kasuya1956a, yosida1957a}, we find that the spin orientation can be controlled by applying a supercurrent flowing through the SC. In the presence of the supercurrent, the quasiparticle bands become asymmetric in momentum space due to the broken parity symmetry, which also modulates the RKKY interaction. Further, the resulting sign change in the RKKY interaction causes the preferred spin orientation to be switched between parallell and antiparallel alignments, both by varying the magnitude of the supercurrent as well as its direction, providing two experimental routes to
observe this effect.

\textit{Theory --} We consider a conventional Bardeen-Cooper-Schrieffer (BCS) \cite{bardeen_pr_57} SC with two impurity spins on its surface. For the superconducting part of the Hamilton-operator, the presence of a supercurrent can be modelled by allowing the order parameter to have a phase gradient. Thus, we may write in real space
\begin{align}
    H_\text{SC} = \frac{\Delta_0}{2}\sum_{i\alpha\beta} \e{\ii \vecQ \cdot \vecr_i} (\ii \sigma^y)_{\alpha\beta} c_{i\alpha}^\dag c_{i\beta}^\dag + \text{h.c.}
\end{align}
Here, $\Delta_0$ is the magnitude of the superconducting order parameter, $\vecQ$ quantifies the magnitude and direction of the supercurrent, whereas $c_{i\sigma}^\dag$ are electron creation operators at site $i$ for spin $\sigma$. For $\vecQ=0$, $H_{\text{SC}}$ reduces to the standard BCS Hamiltonian.

The full Hamilton-operator for the superconducting part, which includes a hopping term, takes the form 
\begin{align}
H_0&=
    \frac{1}{2}\sum_{\vk\sigma}\phi_{\vk\sigma}^\dag \begin{pmatrix}
\epsilon_{\vk}&\sigma\Delta_0\\\sigma\Delta_0&-\epsilon_{-\vk-\vecQ}
\end{pmatrix} \phi_{\vk\sigma},
\label{eq:H_k}
\end{align}
after performing a Fourier transformation $c_{i\sigma}^{\dag}=\frac{1}{\sqrt{N}}\sum_{\vk}c_{\vk\sigma}^\dag \e{\ii \vk \cdot \vecr_i}$ where $N$ is the total number of the lattice points. Above, $\epsilon_{\vk} = -2t[\cos(k_xa) + \cos(k_za)]-\mu$ is the dispersion relation, in which $t$ is the hopping parameter, $a$ is the lattice constant, and $\mu$ is the chemical potential. Here a 2D model in the $xz$-plane is chosen for concreteness and $\phi_{\vk\sigma}^\dag = (c_{\vk\sigma}^\dag \quad c_{-\vk-\vecQ,-\sigma})$ is the fermion basis. The two pairs of energy eigenvalues and eigenstates of the matrix in Eq. (\ref{eq:H_k}) are obtained as $E_{\vk}^{+}$ with $(u_{\vk},\sigma v_{\vk})^T$ and $E_{\vk}^{-}$ with $(-\sigma v_{\vk},u_{\vk})^T$, in which $E_{\vk}^{\pm}=\frac{1}{2}(\epsilon_{\vk}-\epsilon_{-\vk-\vecQ}\pm\sqrt{(\epsilon_{\vk}+\epsilon_{-\vk-\vecQ})^2+4\Delta_0^2})$ and
\begin{align}
u_{\vk} (v_{\vk})=\sqrt{\frac{1}{2}(1 +(-) \frac{\epsilon_{\vk}+\epsilon_{-\vk-\vecQ}}{\sqrt{(\epsilon_{\vk}+\epsilon_{-\vk-\vecQ})^2+4\Delta_0^2}})}.
\end{align}
Based on the eigenpairs, the Hamiltonian is diagonalized as
\begin{equation}
    H_0=\frac{1}{2}\sum_{\vk\sigma}(E_{\vk}^{+}-E_{-\vk-\vecQ}^{-})\gamma_{\vk\sigma}^\dag \gamma_{\vk\sigma},
\end{equation}
where the operators satisfy
\begin{equation}
 \phi_{\vk\sigma} =\begin{pmatrix}
    c_{\vk\sigma}\\ c_{-\vk-\vecQ,-\sigma}^\dag\end{pmatrix}= \begin{pmatrix}
        u_{\vk}&-\sigma v_{\vk}\\\sigma v_{\vk}& u_{\vk}
    \end{pmatrix}\begin{pmatrix}
    \gamma_{\vk\sigma}\\\gamma_{-\vk-\vecQ,-\sigma}^\dag\end{pmatrix}. \label{eq:c_to_gamma}
\end{equation}

\begin{figure*}[t!]
    \centering
    \includegraphics[width = 0.99\textwidth]{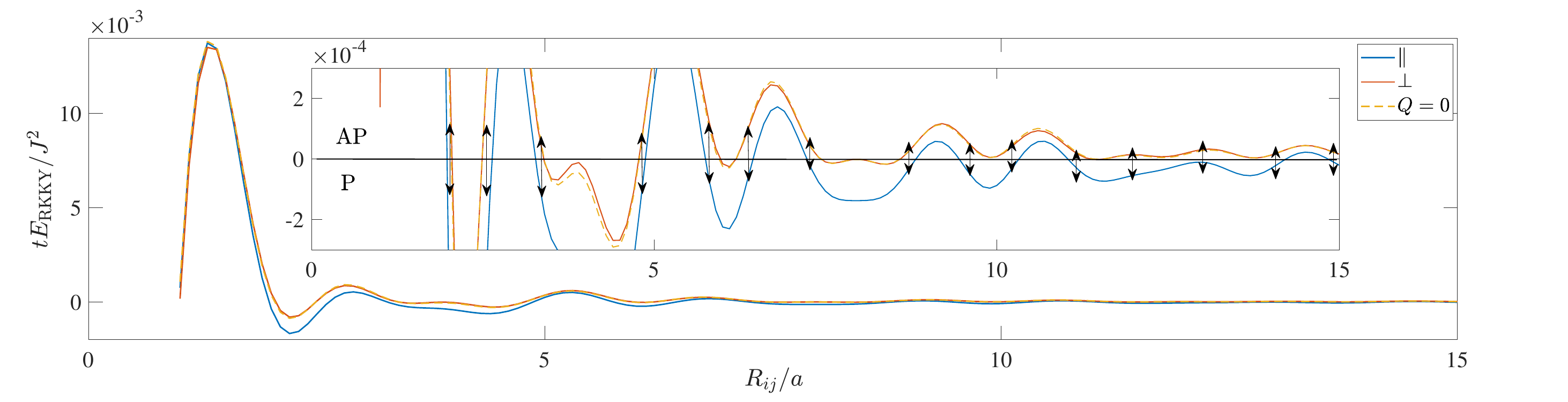}
    \caption{(Color online) Normalized RKKY interaction between two impurity spins on top of a current-carrying superconductor with $Qa=0.1$. The inset shows a zoom-in of the main plot and the horizontal black line is a guide to the eye for where the RKKY interaction changes from P to AP. The direction of the supercurrent is along the impurity separation distance for $\parallel$ and perpendicular to it for $\perp$. $E_\text{RKKY}>0$ favors an AP alignment of the spins, whereas $E_\text{RKKY}<0$ favors a P alignment. The arrows show the preferred spin alignment is altered by either turning the supercurrent on and off or by changing its direction between $\parallel$ and $\perp$. Since the supercurrent magnitude is small for $Qa=0.1$, the $Q=0$ and $\perp$ cases essentially coincide. We consider $\mu/t = -1.8, \Delta_0/t=0.1, k_BT/t=0.01, Q=0.1/a$ and $N=10^4$ sites.}  
    \label{fig:Q1}
    \end{figure*}

To model the impurity spins interacting with the SC, we consider a Hamilton-operator which is treated as a perturbation: $\Delta H = J\sum_{j} \boldsymbol{S}_j \cdot \boldsymbol{s}_j,$
in which $J$ is the strength of the interaction between the impurity classical spin $\boldsymbol{S}_j$ and the conduction electron spin $\boldsymbol{s}_j = \sum_{\alpha\beta}c_{j\alpha}^\dag \boldsymbol{\sigma}_{\alpha\beta}c_{j\beta}$ where $\boldsymbol{\sigma}$ denotes the Pauli matrix vector. We set $S\equiv |\boldsymbol{S}_j|=1$, meaning that the magnitude of the impurity spin is absorbed into the coupling constant $J$. 
After Fourier-transforming and expressing the $c$-operators in terms of $\gamma$-operators described by Eq. (\ref{eq:c_to_gamma}), we obtain
\begin{align}
    \Delta H &= \sum_{\vk\vk{'}\alpha\beta}T_{\vk\vk{'}\alpha\beta}[u_{\vk}^{*}u_{\vk^{'}}\gamma_{\vk\alpha}^\dag \gamma_{\vk^{'}\beta} - \beta u_{\vk}^{*}v_{\vk^{'}}\gamma_{\vk\alpha}^\dag\gamma_{-\vk^{'}-\vecQ,-\beta}^\dag \notag\\
    &-\alpha v_{\vk}^{*}u_{\vk^{'}}\gamma_{-\vk-\vecQ,-\alpha}\gamma_{\vk^{'}\beta}+\alpha\beta v_{\vk}^{*}v_{\vk^{'}}\gamma_{-\vk-\vecQ,-\alpha}\gamma_{-\vk^{'}-\vecQ,-\beta}],
    \label{eq:Delta_H}
\end{align}
in which $T_{\vk\vk^{'}\alpha\beta}=\sum_j \frac{J}{N}e^{i(\vk-\vk^{'})\cdot \boldsymbol{r}_j}\boldsymbol{S}_j\cdot\boldsymbol{\sigma}_{\alpha\beta}$ is defined.

We now perform a Schrieffer-Wolff transformation to obtain the RKKY interaction between the impurity spins, mediated by the SC. This is in essence a second order perturbation theory for $\Delta H$ achieved by applying a
canonical transformation $H_{\text{eff}}
= e^{\eta S} H e^{-\eta S}$ for $H = H_0 + \Delta H$. Subsequently, one identifies $\eta S$ so that it satisfies $\Delta H +[\eta S,H_0]=0$ which projects out the first order effect of the perturbation, which does not generate any interaction between the impurity spins. This
gives rise to the effective Hamiltonian
\begin{equation}
H_\text{eff}=H_0+\frac{1}{2}[\eta S,\Delta H],
\label{eq:H_eff}
\end{equation}
in which one can express $\eta S$ with the same operators as in Eq. (\ref{eq:Delta_H}): $\eta S =\sum_{\vk\vk{'}\alpha\beta}$$[A_{\vk\vk{'}\alpha\beta}\gamma_{\vk\alpha}^\dag \gamma_{\vk^{'}\beta} + B_{\vk\vk{'}\alpha\beta}\gamma_{\vk\alpha}^\dag\gamma_{-\vk^{'}-\vecQ,-\beta}^\dag
    + C_{\vk\vk{'}\alpha\beta}\gamma_{-\vk-\vecQ,-\alpha}\gamma_{\vk^{'}\beta}+$$D_{\vk\vk{'}\alpha\beta}\gamma_{-\vk-\vecQ,-\alpha}\gamma_{-\vk^{'}-\vecQ,-\beta}].$
The coefficients are consequently identified as 
\begin{align}
    A_{\vk\vk{'}\alpha\beta}&=-\frac{2u_{\vk}^{*}u_{\vk^{'}}T_{\vk\vk{'}\alpha\beta}}{E_{\vk^{'}\beta}^{+}-E_{-\vk^{'}-\vecQ,-\beta}^{-}-E_{\vk\alpha}^{+}+E_{-\vk-\vecQ,-\alpha}^{-}},\notag\\
     B_{\vk\vk{'}\alpha\beta}&=-\frac{2\beta u_{\vk}^{*}v_{\vk^{'}}T_{\vk\vk{'}\alpha\beta}}{E_{\vk\alpha}^{+}-E_{-\vk-\vecQ,-\alpha}^{-}+E_{-\vk^{'}-\vecQ,-\beta}^{+}-E_{\vk^{'}\beta}^{-}},\notag\\
     C_{\vk\vk{'}\alpha\beta}&=\frac{2\alpha v_{\vk}^{*}u_{\vk^{'}}T_{\vk\vk{'}\alpha\beta}}{E_{\vk^{'}\beta}^{+}-E_{-\vk^{'}-\vecQ,-\beta}^{-}+E_{-\vk-\vecQ,-\alpha}^{+}-E_{\vk\alpha}^{-}},\notag\\D_{\vk\vk{'}\alpha\beta}&=\frac{2\alpha\beta v_{\vk}^{*}v_{\vk^{'}}T_{\vk\vk{'}\alpha\beta}}{E_{-\vk^{'}-\vecQ,-\beta}^{+}-E_{\vk^{'}\beta}^{-}-E_{-\vk-\vecQ,-\alpha}^{+}+E_{\vk\alpha}^{-}}.
\end{align}
Given $\eta S$, the expectation value of the effective Hamiltonian given by Eq. (\ref{eq:H_eff}) may now be evaluated to obtain the RKKY interaction. 

\textit{Results and discussion --} Defining $S_j^{\alpha\beta} \equiv \boldsymbol{S}_j\cdot\boldsymbol{\sigma}_{\alpha\beta}$ and using $\Sigma_{\alpha\beta}S_j^{\alpha\beta}S_i^{\beta\alpha}=2\boldsymbol{S}_i\cdot \boldsymbol{S}_j$ and  $\Sigma_{\alpha\beta}\alpha\beta S_j^{\alpha\beta}S_i^{-\alpha,-\beta}=-2\boldsymbol{S}_i\cdot \boldsymbol{S}_j$, we obtain the expectation value 
\begin{equation}\label{eq:Heff}
  \langle H_{\text{eff}}\rangle = E_0 + \Sigma_{ij}E_{\text{RKKY}}\boldsymbol{S}_i\cdot \boldsymbol{S}_j, 
\end{equation}
in which $E_0$ is a constant and $E_\text{RKKY}$ describes the RKKY interaction. The sum $\sum_{ij}$ in Eq. (\ref{eq:Heff}) is over the two impurity spins. Applying $u_{\vk}=u_{-\vk-\vecQ}$, $v_{\vk}=v_{-\vk-\vecQ}$ and $E_{\vk}^{\pm}=-E_{-\vk-\vecQ}^{\mp}$, the RKKY interaction can after lengthy calculations be expressed via the quantities
\begin{align}
    F_1(\vk,\vk') &= (|u_{\vk} u_{\vk^{'}}|^2 + u_{\vk}^{*}u_{\vk^{'}}v_{\vk}^{*}v_{\vk^{'}})\frac{n(E_{\vk}^{+})-n(E_{\vk^{'}}^{+})}{E_{\vk^{'}}^{+}-E_{\vk}^{+}},\notag\\
    F_2(\vk,\vk') &= (-|u_{\vk} v_{\vk^{'}}|^2 + u_{\vk}^{*}u_{\vk^{'}}v_{\vk}^{*}v_{\vk^{'}})\frac{n(E_{\vk}^{+})+n(E_{-\vk^{'}-\vecQ}^{+})-1}{E_{\vk}^{+}+E_{-\vk^{'}-\vecQ}^{+}},\notag\\
    F_3(\vk,\vk') &= (u_{\vk}^{*}u_{\vk^{'}}v_{\vk}^{*}v_{\vk^{'}}-|u_{\vk^{'}} v_{\vk}|^2)\frac{n(E_{\vk^{'}}^{+})+n(E_{-\vk-\vecQ}^{+})-1}{E_{\vk^{'}}^{+}+E_{-\vk-\vecQ}^{+}},\notag\\
    F_4(\vk,\vk') &= (u_{\vk}^{*}u_{\vk^{'}}v_{\vk}^{*}v_{\vk^{'}}+|v_{\vk} v_{\vk^{'}}|^2)\frac{n(E_{-\vk-\vecQ}^{+})-n(E_{-\vk^{'}-\vecQ}^{+})}{E_{-\vk^{'}-\vecQ}^{+}-E_{-\vk-\vecQ}^{+}},
\end{align}
in the following form
\begin{align}\label{eq:RKKY}
   E_\text{RKKY} &= -(\frac{J}{N})^2\Sigma_{\vk\vk^{'}}
e^{i(\vk-\vk^{'})\cdot\boldsymbol{R}_{ij}}\Big[F_1(\vk,\vk') + F_2(\vk,\vk') \notag\\
&+ F_3(\vk,\vk') + F_4(\vk,\vk')\Big]
\end{align}
where $\boldsymbol{R}_{ij}=\boldsymbol{r}_{j}-\boldsymbol{r}_{i}$ and $n(E)=(1+e^{\beta E})^{-1}$ denotes the Fermi-Dirac distribution at energy $E$ with $\beta=1/k_B T$. The above expression can be further simplified since $E_\text{RKKY}$ is real, and thus the exponential prefactor can be replaced with its corresponding cosine component. Subsequently, one observes that the contribution from $F_2$ is the same as $F_3$, which can be seen by renaming indices $\vk \leftrightarrow \vk'$ and using that $u,v$ are real. 

\begin{figure*}[t!]
    \centering

    \includegraphics[width = 0.99\textwidth]{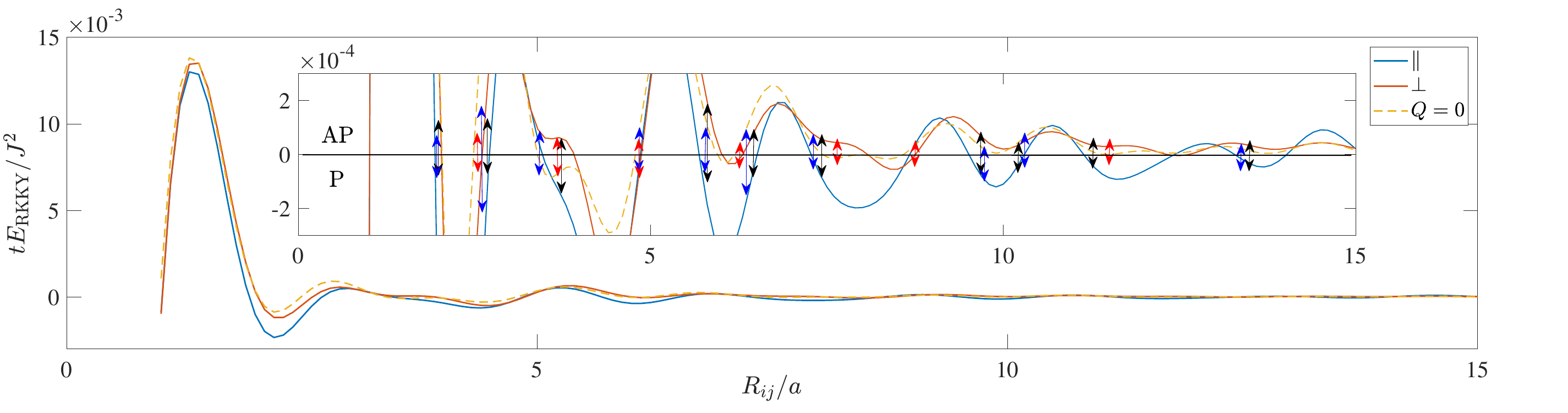}
    \caption{(Color online) 
    Same as in Fig. \ref{fig:Q1}, but for $Qa=0.2$. Since the supercurrent is now larger than in Fig. \ref{fig:Q1}, additional spin switching is enabled. Namely, the preferred spin alignment is switched by either turning the supercurrent on and off in the $\parallel$ direction (blue arrows), on and off in the $\perp$ direction (red arrows), or changing the direction of the supercurrent
 between $\parallel$ and $\perp$ (black arrows). We consider $\mu/t = -1.8, \Delta_0/t=0.1, k_BT/t=0.01, Q=0.2/a$ and $N=10^4$ sites.}
    \label{fig:Q2}
\end{figure*}

\begin{figure}[b!]
    \centering
    \includegraphics[width = 0.47\textwidth]{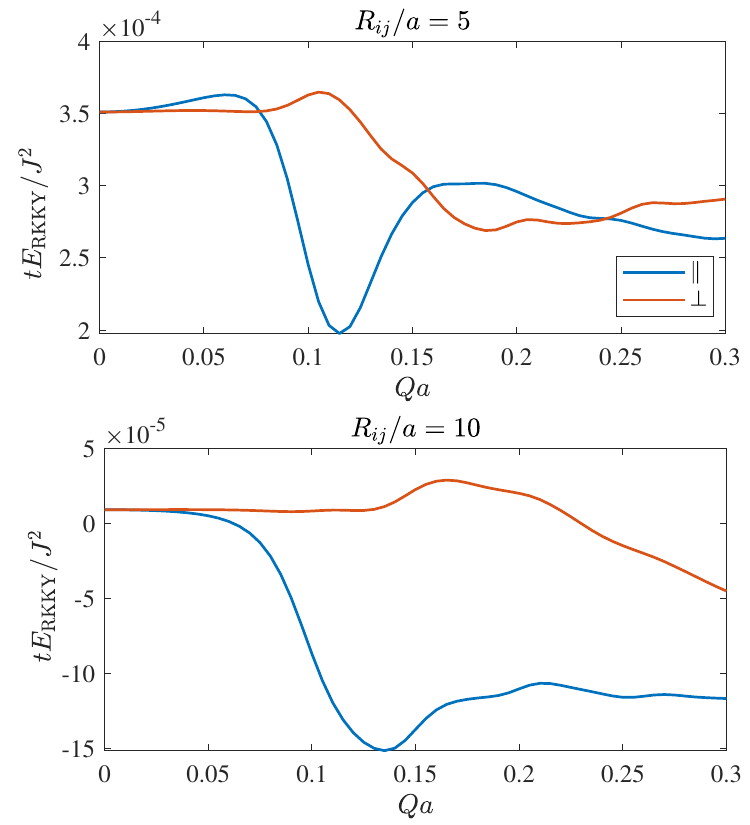}
    \caption{(Color online) Normalized RKKY interaction as a function of supercurrent magnitude. We consider two separation distances in the top and bottom panels and consider both a supercurrent flow along $(\parallel)$ the separation distance vector and perpendicular $(\perp)$ to it. We set $\mu/t = -1.8, \Delta_0/t=0.1, k_BT/t=0.01$, and $N=10^4$ sites. } 
    \label{fig:Rij}
\end{figure}

For $\vecQ=0$, we regain the results studied previously in the literature for RKKY interaction in SCs  \cite{alekseevskii_zetf_77, kochelaev_zetf_79, khusainov_zetf_96, aristov_zpb_97} in the form of an additional antiferromagnetic, exponentially decaying term that appears in $E_\text{RKKY}$ along with the usual rapidly oscillating interaction.
Eq. (\ref{eq:RKKY}) can then be numerically evaluated to determine the effect of a supercurrent on the spin-spin interaction. To estimate a reasonable magnitude for the momentum $Q=|\vecQ|$ of the Cooper pairs, we note that the critical supercurrent that a SC can sustain is provided by $Q\xi \simeq 1$ \cite{tinkham_book} where $\xi = \hbar v_F/(\pi\Delta_0)$ is the coherence length. An analytical estimate for $Q$ can be given for a simple 1D model. The Fermi velocity in our lattice model is defined via $v_F = \frac{1}{\hbar} (d\epsilon_k/dk)|_{k=k_F}$
where $k_F$ is obtained as the momentum where $\epsilon_k = 0$. To maximize the value of $Q$ (in order to have a supercurrent which can strongly influence the RKKY interaction), one ideally needs a SC with as small $\xi$ as possible. High-$T_c$ superconductors can have $\xi \simeq 3a$, allowing $Q \simeq 0.3/a$. Subsequently, $\mu$ and $\Delta_0$ should be chosen to get $\xi \simeq 3a$. Choosing $\mu=-1.8t$, one finds from $-2t\cos (ka)-\mu=0$ that $k_F \simeq 0.5/a$, which gives $v_F \simeq at/\hbar$. Then, for $\Delta_0/t = 0.1$, we can achieve $\xi \simeq 3a$ which gives the upper limit $Q \simeq 0.3/a$. Similar parameters were used in Ref. \cite{takashima_prb_18}.

The RKKY interaction results are shown in Fig. \ref{fig:Q1} for $Q=0.1/a$. We show results both for zero supercurrent $(Q=0)$, supercurrent flowing parallell ($\parallel$) and perpendicular ($\perp$) to the separation vector $\vecR_{ij}$ of the two spin impurities. Here we fix $\vecQ$ along $\vecx$ and consider $\vecR_{ij}$ along $\vecx$ ($\vecz$) to cover the $\parallel$ ($\perp$) configuration. The figure demonstrates that the RKKY interaction changes its sign within several separation distance regimes by tuning the magnitude and direction of the supercurrent.  Since $E_\text{RKKY}<0$ causes a parallel (P) alignment of the two spin impurities while $E_\text{RKKY}>0$ supports an antiparallel (AP) orientation, the sign change thus induces spin switching between the P and AP states. In addition, the RKKY curves are almost the same for the $Q=0$ and $\perp$ cases. This can be explained by the energy dispersion symmetry breaking induced by the supercurrent, which is the strongest 
for the quasiparticles mediating the RKKY interaction when $\vecQ \parallel \vecR_{ij}$ and negligible for the perpendicular case when $Q$ is small. In Fig. \ref{fig:Q1}, the black arrows denote spin switching achieved by changing the direction of supercurrent flow (between $\vecQ \parallel \vecR_{ij}$ and $\vecQ \perp \vecR_{ij}$). The black arrows also indicate switching caused by turning the supercurrent on and off (between $\vecQ \parallel \vecR_{ij}$ and $\vecQ=0$) since the $Q=0$ and $\perp$ cases essentially coincide due to the small value of $Q$. It is clear from the arrows in the figure that the presence of supercurrent gives rise to ample opportunities for spin switching at several separation distances. Note that for each arrow, the switch occurs in a finite interval centered around the position of the arrow and not just exactly at the location of the arrow, making the switching effect more accessible.

We also show results in Fig. \ref{fig:Q2} for a slightly larger value of the supercurrent, $Q=0.2/a$, demonstrating the robustness of the effect and that there exists an abundance of possible switching effects by either turning of the supercurrent or by changing its direction. As $Q$ increases, compared with $Q=0.1/a$ in Fig. \ref{fig:Q1}, the difference between the $Q=0$ and $\perp$ cases becomes distinguishable and the additional spin switching between them becomes possible, as the red arrows show.

Finally, we plot the RKKY interaction energy at a fixed lattice site as a function of the supercurrent magnitude $Q$ in Fig. \ref{fig:Rij} for two site choices. The supercurrent flow starts modifying $E_\text{RKKY}$ at much smaller values of $Q$ when it flows along $\vecR_{ij}$ compared to when it flows perpendicular to it. The physical mechanism behind this is the directional dependence of the asymmetry in the quasiparticle bands $E_{\vk}$ created by $\vecQ$. As mentioned before, 
the asymmetry is strongest for particles moving between the impurity spins when $\vecQ \parallel \vecR_{ij}$, which are precisely the ones contributing the most to the RKKY interaction. The lower panel of Fig. \ref{fig:Rij} shows that at a fixed separation distance $R_{ij}=|\vecR_{ij}|$, modulating the supercurrent magnitude $Q$ can cause the preferred spin orientation to switch between P ($E_\text{RKKY} <0$) and AP ($E_{\text{RKKY}} > 0$), which is consistent with the switching results observed in Figs. (\ref{fig:Q1},\ref{fig:Q2}). 

We also give an estimate for the effect of the supercurrent Oersted field acting on the impurity spins via a Zeeman-effect, and show that it is negligible compared to the RKKY interaction. Considering a thin superconducting film of thickness $d$ with a critical current density $J_c = 10^7$ A/cm$^2$, the field at the surface can be approximated as $B = \mu_0 J_c d/2$ at the critical supercurrent strength. For $d=15$ nm, this gives $B \simeq 10^{-3}$ T, corresponding to a very small Zeeman-coupling $E_Z \simeq 10^{-5}$ meV at about half of the critical current density. This can be compared to $tE_\text{RKKY}/J^2$ in our plots, which is typically of order $10^{-4}$ at a separation distance of several lattice sites. Using a weak impurity spin coupling $J=0.05t<\Delta_0$, as appropriate for the perturbative approach employed here, we get for $t=500$ meV that $E_\text{RKKY} \simeq 10^{-4}$ meV which is $\gg E_Z$. Although this is a rough estimate, we note that larger couplings $J$, outside the regime of our approach, between the impurity and conduction electron spins are accessible experimentally \cite{yazdani_science_97}. This will make the RKKY interaction even larger, in particular compared to the Oersted-field effect. A thinner SC film decreases the Oersted field further. The effect of supercurrent flow in the strong-coupling regime could be an interesting topic for future studies where in-gap Yu-Shiba-Rusinov states \cite{yu_aps_65, shiba_ptp_68, rusinov_zetf_69} are expected to have a more prominent role. The main conclusion of this work, being the tunability of the RKKY interaction and thus the possibility to switch the ground state spin configuration, is expected to hold also in this case.

\textit{Concluding remarks --}  Our proposed system setup should be experimentally feasible. In Ref. \cite{kuster_natcom_21}, the RKKY interaction between Cr impurity spins coupled to a SC was studied using scanning tunneling spectroscopy. All that is required in addition to observe the supercurrent-induced spin switching is the application of a current bias to the SC. We hope that the present work will stimulate the anticipated experimental realization of supercurrent-induced spin switching.

\textit{Acknowledgments --}
This work was supported by the Research Council of Norway through Grant No. 323766 and its Centres of Excellence funding scheme Grant No. 262633 “QuSpin.” Support from Sigma2 - the National Infrastructure for High-Performance Computing and Data Storage in Norway, project NN9577K, is acknowledged.

\bibliography{masterref}

\end{document}